\documentstyle[12pt,a4,epsf]{article}
\begin{document} 
\begin{center}
\begin{Large}
QUARKONIUM IN THE WILSON LOOP FORMALISM: NEW PERSPECTIVES
\end{Large}
\vspace{0.4cm}\\
N. Brambilla \footnote{Plenary talk presented at  ``Nuclear and Particle 
physics with CEBAF at Jefferson Lab'', Dubrovnik, November 3-10, 1998.}$^a$ and A. Vairo$^b$ \\
{\small\em $^a$ Institut f\"ur Theoretische Physik, Universit\"at Wien}\\ 
{\small\em    Boltzmanngasse 5, A-1090 Vienna, Austria}\\
{\small\em $^b$ Institut f\"ur Hochenergiephysik, \"Osterreichische Akademie der Wissenschaften}\\
{\small\em  Nikolsdorfergasse 18, A-1050 Vienna, Austria}
\end{center}

\begin{abstract}
We review the present knowledge on the heavy quark interaction. The framework is the NRQCD effective theory 
and the interaction is expressed in terms of Wilson loop chromoelectric and chromomagnetic insertions.
\end{abstract}

\section{Introduction}

The experimental evidence that for heavy quark bound states like $b\bar{b}$, $c\bar{c}$, ... all the 
splittings are considerably less than the masses suggests that all the dynamical energy scales 
of these systems are small with respect to the quark masses. As a consequence the quark velocities $v$ 
are small and these systems can be considered as non relativistic. The hierarchy of the scales is then 
the typical one of a non relativistic system. Called $m$ the mass of the heavy quark, 
the quark momenta scale like $m v$ and the quark energies like $m v^2$. Trickier is the 
situation in the gluon sector, but the binding interaction is essentially characterized by the 
same energy scale distribution. Therefore the dominant gluon interaction among heavy quarks 
appears ``instantaneous''. A potential picture should hold, at least in first approximation, 
and the energy levels can be obtained by solving the corresponding Schr\"odinger equation.
In particular for infinitely heavy quarks the spin splittings vanish and the system 
can be described only by means of a single static central potential.

Indeed, potential models have been quite successful \cite{rep} in explaining the hadron spectrum.
However, being  these   phenomenological models,  their  connection with the QCD parameters 
is hidden, the scale at which they are defined is not clear and they cannot be systematically improved.  
In this paper we outline the kind of {\it rigorous description of quarkonium} that  can be obtained 
from QCD at the present. 

Several perturbative evaluations of the quarkonium potential have been performed in the last twenty years 
\cite{buchmuller} and recently the complete $\alpha_{\rm s}^2$ corrections to the levels have been 
calculated \cite{yndurain}. The main difficulty in these calculations is connected with the inclusion 
of nonperturbative contributions which not only set a bound on the reachable precision, 
but turn out to be relevant for quantitative predictions even on the bottomonium spectrum. 
A typical approach is to consider all the nonperturbative physics to be encoded 
into few local condensates. This corresponds to assume that the typical length associated, for instance,  
with the nonperturbative gluodynamics is larger than any other length scale of the system and can 
be put equal to infinity. In a pure analytic calculation this statement is also dictated by the 
necessity to have a small number of nonperturbative parameters and maintain predictability.  
Even for the ground state this assumption is doubtful, but surely it does not hold for large 
radii quarkonia (i.e. excited heavy mesons) where the nonperturbative gluonic length 
cannot be considered large with respect to  the size of the bound state. Therefore a proper  
treatment of nonperturbative effects includes {\it non-local} condensates. Another way to say the 
same thing is to notice that the contribution to the levels associated with the local gluon condensate 
is proportional to $n^6 \langle \alpha_{\rm s} F^2(0)\rangle$. By increasing the principal 
quantum number $n$ beyond the ground state, it grows very soon out of control. 

A way of including  nonperturbative effects, as they are, in the evaluation 
of the quarkonium spectrum is lattice QCD. This technique is becoming more and more 
successful and in the near future it is expected to be the only competitive one (see for 
instance \cite{davies} and references therein). There are different ways in which 
lattice QCD calculations can be performed. We mention lattice NRQCD \cite{lepage} where 
the quarkonium spectrum is directly evaluated on the lattice by means of an effective 
action derived from QCD by an expansion in the quark velocities.  
Very close is the approach we will discuss in the following where 
the QCD Lagrangian is replaced by its effective non relativistic formulation 
{\it before} doing any lattice evaluation. All the non perturbative physics 
is encoded in this way in the Wilson loop made up by the quark trajectories 
and in field strength insertions on it. These (non local) objects are then evaluated 
on the lattice. We call this the {\it Wilson loop approach}. The advantage is that in this way 
our expressions are safer to handle for lattice purposes, since less affected by finite size 
effects, and easier to treat also for analytic purposes like the implementation of vacuum models. 
In particular we get an expression for the heavy quark-antiquark potential. 
In the limit where the insertions of two field strengths on the Wilson loop can be approximated 
by a local condensate, one gets back the ``improved'' perturbative expression discussed previously.  

The (Wegner)-Wilson loop formalism has a long story. It was first suggested by K. Wilson \cite{wilson} 
that the object called after him Wilson loop would be the relevant one in order to describe 
confinement. The strong coupling expansion suggested an area law behaviour further confirmed 
by lattice simulations, which were born with this pioneering work. In \cite{brown} it was  
shown how to relate rigorously the static Wilson loop with the static quark-antiquark potential. 
On the same line few years later also spin dependent corrections to the potentials where expressed 
in terms of Wilson loop and chromoelectric and chromomagnetic field insertions \cite{eichten}. 
In particular it was proven that a nonperturbative behaviour in the static potential must give rise  
to nonperturbative spin dependent corrections as well. Non spin dependent corrections were treated 
in the same framework some time later \cite{consoli}. 
This was the situation pictured at the beginning of the '90 in \cite{rep}. Several problems were 
still open. No relativistic formulation in terms of the Wilson loop was available. 
There was an apparent mismatch between the Eichten--Feinberg--Gromes expression for the spin dependent 
sector of the potential (analytic in the quark masses) and the perturbative one loop estimate (containing 
logarithms of the quark masses). The inclusion of non potential terms was not understood. 
While not all these problems have been solved, some remarkable progress has been achieved 
in the last years. Attempts in the direction of a relativistic formulation have been done in \cite{fs}. 
In \cite{chen} it was shown that, by performing properly the matching between the effective theory and QCD, 
Wilson coefficients carrying logarithms of the quark masses appear in the Eichten--Feinberg--Gromes 
expression for the potential. The inclusion of non potential contributions in this framework is
now  in progress \cite{inprep} and the way to precision calculations in quarkonium seems finally to be open. 
Here we only mention that this goal is met by  a new effective theory  built from NRQCD, where 
explicitly potential and non potential terms have been separated \cite{pnrqcd}. Finally, all the 
Wilson loop averages relevant for the potential have been calculated on the lattice and the bottomonium 
and charmonium spectra have been calculated with good agreement with the data \cite{bali,brambilla}. 

In the meantime there have been several progress in the building of a non relativistic 
effective theory from QCD, mainly due to the success of Heavy Quark Effective Theory 
in describing heavy-light systems \cite{hqet}. Wilson coefficients have been calculated at higher 
order and the role played by reparameterization invariance has been better understood.

Here, we summarize the present level of understanding of the heavy quark potential 
which contributes to the energy levels of quarkonium at the order $v^4$. 
The framework is NRQCD, the tool the Wilson approach.

\section{NRQCD}\label{secnrqcd}

In order to define an effective theory we typically need three ingredients: an effective Lagrangian,  
a regularization scheme and therefore a matching scale and a power counting set of rules. 
The effective theory we will use is NRQCD \cite{lepage}. Let us discuss its key ingredients. 

The NRQCD Lagrangian is obtained from QCD by expanding with respect to the heavy quark masses. 
The matching with QCD is performed like in Heavy Quark Effective Theory (HQET) \cite{manohar,pineda}. 
We emphasize that in order to build up the effective Lagrangian from the QCD Lagrangian,  
we have to ignore the specific dynamical problem we are dealing with and expand with respect to  
the heavy quark masses which are explicit parameters of QCD.   Typically the effective Lagrangian 
turns out to be the sum of a pure gauge part $L_{\rm g}$ plus two, four, ... fermion terms 
($L_{\rm 2f}$, $L_{\rm 4f}$, ...). Since we are interested in two-body bound states we will 
take into account only two and four fermion terms (terms involving more fermions 
will contribute only in intermediate states). Therefore the NRQCD Lagrangian 
obtained from QCD by expanding with respect of the mass $m_1$ of a heavy quark and 
the mass $m_2$ of a heavy antiquark is given by \cite{manohar,pineda}
\begin{equation}
L = L_{\rm 2f} + L_{\rm 4f} + L_{\rm g},
\label{nrqcd}
\end{equation}
where at order $1/m^2$ and up to field redefinitions 
\begin{eqnarray}
L_{\rm 2f} &=& Q_1^\dagger \left( iD_0 + c^{(1)}_2 {{\bf D}^2\over 2 m_1} + 
c^{(1)}_4 {{\bf D}^4\over 8 m_1^3} + c^{(1)}_F g {  \mbox{ {\boldmath $\sigma$}} \cdot {\bf B} \over 2 m_1} 
+ c^{(1)}_D g { {\bf D} \cdot {\bf E} - {\bf E} \cdot {\bf D} \over 8 m_1^2} \right. 
\nonumber\\
& & \qquad \left. + i c^{(1)}_S g { \mbox{ {\boldmath $\sigma$}}  
\cdot ({\bf D} \times {\bf E} - {\bf E} \times {\bf D})
\over 8 m_1^2} \right)Q_1 + O\left( {1\over m_1^3}\right)
\nonumber\\
&+& Q_2^\dagger\left(-iD_0 + c^{(2)}_2 {{\bf D}^2\over 2 m_2} + 
c^{(2)}_4 {{\bf D}^4\over 8 m_2^3} + c^{(2)}_F g {  \mbox{ {\boldmath $\sigma$}} \cdot {\bf B} \over 2 m_2} 
- c^{(2)}_D g { {\bf D} \cdot {\bf E} - {\bf E} \cdot {\bf D} \over 8 m_2^2} \right. 
\nonumber\\
& & \qquad \left. - i c^{(2)}_S g {\mbox{ {\boldmath $\sigma$}}  
\cdot ({\bf D} \times {\bf E} - {\bf E} \times {\bf D})
\over 8 m_2^2} \right)Q_2 + O\left( {1\over m_2^3}\right), 
\label{nrqcd2f}
\end{eqnarray}
$Q_j$ are the heavy quark fields and the covariant derivative is defined as $D_\mu = \partial_\mu + igA_\mu^aT^a$. 
The explicit form of $L_{\rm 4f}$ is given in \cite{pineda}. $L_{\rm g}$ is the SU(3) Yang--Mills Lagrangian 
modified in order to give rise to an effective  $\alpha_{\rm s}$ running with 2 (heavy) flavors less. 

The effective Lagrangian (\ref{nrqcd}) is not renormalizable. Therefore, it is necessary to 
regularize it. In a given regularization scheme the reproduction of the correct 
ultraviolet regime of QCD is obtained by means of the Wilson coefficients. 
The effective Lagrangian is complete only once these coefficients are given. 
 The Wilson coefficients are evaluated at a matching scale where perturbation 
theory still holds. They encode the ultraviolet regime of QCD 
up to a given scale $\mu$ order by order in the coupling constant $\alpha_{\rm s}$. 
Renormalization group (RG) transformations should be used in order to resum leading log contributions 
($\sim \ln m/\mu$). The matching coefficients are known in the literature at different level of precision 
(i.e. at different order in the coupling constant), see \cite{manohar,pineda,let}
Here we remember only that reparameterization invariance \cite{manohar} (i.e. the invariance of 
the effective Lagrangian with respect a variation of $v$) fixes $c^{(j)}_2=c^{(j)}_4=1$.   

Heavy quark bound states are characterized by a dynamical dimensionless parameter, the quark velocity $v$. 
As explained in the introduction, this parameter is small. From phenomenological potential models \cite{quigg} 
and from lattice studies \cite{bali} we get the usually accepted values of $\langle v^2_b \rangle \sim 0.07$ 
for the bottomonium system and $\langle v^2_c \rangle \sim 0.24$ for the charmonium system. 
This allows the ordering of the energy scales of the problem, $m$, $m v$ and $m v^2$. 
The first scale has been explicitly integrated out in the Lagrangian (\ref{nrqcd}) and, as discussed above, 
the contributions coming from it are carried order by order in $\alpha_{\rm s}$ by the Wilson coefficients.  
The last two scales, sometimes known with the name ``soft'' and ``ultrasoft'' respectively, are still mixed up. 
The relation between $v$ and the QCD parameters is in general unknown. It is expected to be 
the result of perturbative and nonperturbative effects. For infinitely heavy quarks $v$ coincides 
with $\alpha_{\rm s}$ like in QED where, for instance in the hydrogen atom, the hyperfine structure 
constant $\alpha$ is equal to the electron velocity. The main point is that the existence of this small 
dynamical parameter enables us to set up power counting rules for the operators in the effective 
Lagrangian.  For the sake of simplicity we reproduce here the rough power counting argument 
of \cite{lepage}, keeping in mind that an exact power counting cannot be done until soft and ultrasoft 
degrees of freedom have not been completely disentangled. Noticing that the number operator 
for heavy quarks, $\displaystyle \int d^3x \,Q^\dagger(x) Q(x)$, does not depend on $v$,  
one gets $Q \sim (mv)^{3/2}$. The kinetic energy $\displaystyle \int d^3x\, Q^\dagger(x) {{\bf D}^2\over 2m}Q(x)$ 
is by definition of order $mv^2$ and therefore ${\bf D} \sim mv$ (i.e. the contribution of ${\bf D}$ 
to the levels is of order $mv$). Finally, using the equation of motion one gets the other counting 
rules $D_0 \sim m v^2$,  $gA_0 \sim m v^2$ $g{\bf A} \sim m v^3$, $gE\sim m^2 v^3$ and $gB \sim m^2 v^4$. 
With respect to these rules the Lagrangian of Eq. (\ref{nrqcd2f}) is accurate up to 
order $O(v^4)$ of the levels (or up to $O(v^2)$ with respect to the leading contribution). 
This should guarantee a rough 10\% of accuracy on the absolute value of the levels. 

Concluding this section, we stress that the NRQCD power counting defined above is not the same as that  
one used in HQET. In particular, in HQET the kinetic energy ${\bf D}^2/2 m$ is suppressed by a factor 
$\Lambda_{\rm QCD}/m$ with respect to the operator $D_0$ while in NRQCD the two operators are of the same order. 
As a consequence the heavy quark propagator contains in NRQCD a kinetic part which is absent in HQET where 
the heavy quark propagator is static. In a very general way one can say that these differences are due to the 
fact that, even if the effective Lagrangian is essentially the same, the physical problem is different 
and it is the physical problem which defines the counting rules.

\section{The Wilson loop formalism}\label{secwilson}

The next step is to derive the heavy quark interaction in the so-called Wilson loop 
formalism. In this context the use of the effective Lagrangian (\ref{nrqcd2f}) (with tree level matching) 
was first suggested by L. S. Brown and W. I. Weisberger, later by E. Eichten and F. Feinberg 
in \cite{brown,eichten}. A one loop RG improved calculation was done in \cite{chen}. In those papers an 
expansion is performed around the static solution.  
Here, we adopt the approach of  \cite{consoli}
 where the kinetic energy was kept during all the calculations.

The 4-point gauge invariant Green function $G$ associated with the Lagrangian (\ref{nrqcd}) is defined as 
$$
G(x_1,y_1,x_2,y_2) =  \langle 0 \vert  Q_2^\dagger(x_2)  \phi(x_2,x_1) Q_1(x_1)  Q_1^\dagger(y_1)  \phi(y_1,y_2) 
Q_2(y_2) \vert 0\rangle, 
$$
where $\phi(x_2,x_1) \equiv \displaystyle\exp\left\{ - ig \int_0^1 ds \, (x_2-x_1)^\mu A_\mu(x_1 + s (x_2-x_1)) \right\}$ 
is a Schwinger line added to select the singlet state contribution. For large time separations the string 
vanishes. After integrating out the heavy quark fields $Q_j$ and $Q_j^\dagger$, 
$G$ can be expressed as a quantum-mechanical path integral over the quark trajectories \cite{fs}:
$$
G  \!=\! \int_{y_1}^{x_1} \!\!\!{\cal D}z_1 {\cal D}p_1\int_{y_2}^{x_2} \!\!\!{\cal D}z_2 {\cal D}p_2\,
\exp\left\{i\!\int_{-T/2}^{T/2} \!\!dt \sum_{j=1}^2 {\bf p}_j\cdot {\bf z}_j 
- {p_j^2\over 2 m_j}  + {p_j^4\over 8 m_j^3} -i \int_{-T/2}^{T/2} \!\! dt \, U\right\}
$$
where $z_j = z_j(t)$ and $y_2^0 = y_1^0 \equiv -T/2$, $x_2^0 = x_1^0 \equiv T/2$. 
The function $U$ describes the heavy quark interaction.  Since the kinetic energy has been properly isolated, 
it is consistent with the counting rules given in the previous section to expand the interaction 
around the static quark trajectories $z_1 = (t,{\bf r})$ and $z_2 = (t,{\bf 0})$.  Moreover, we define 
the heavy quark-antiquark potential as $V = \displaystyle\lim_{T\to\infty} {1\over T} 
\int_{-T/2}^{T/2} \!\!dt \,U$, with the warning that, having soft and ultrasoft degrees of freedom 
not been disentangled in NRQCD, non potential terms could still contribute to some extent to $V$. 

Working out this expression \cite{consoli,fs} we get  
\begin{eqnarray}
& & \!\!\!\!\!\!\!\!\!\!\!
V = V_0 + \hbox{spin indipendent terms} - {1 \over  m_1 m_2}(d_{ss} +  C_F\,d_{vs}) \, \delta^3(r)
\nonumber\\
& & \!\!\!\!\!\!\!\!\! + {1\over 8}\left( {c_D^{(1)} \over  m_1^2} 
+ {c_D^{(2)} \over  m_2^2} \right) (\Delta V_0(r) + \Delta V_{\rm a}^E(r))
+ {1\over 8} \left( {c_F^{(1)} \over  m_1^2} 
+ {c_F^{(2)} \over  m_2^2} \right) \Delta V_{\rm a}^B(r)
\nonumber\\
& & \!\!\!\!\!\!\!\!\! + \left( {{\bf S}^{(1)}\cdot{\bf L}^{(1)}\over m_1^2} +  
{{\bf S}^{(2)}\cdot{\bf L}^{(2)}\over m_2^2} \right)\!
{2 c^+_F V_1^\prime(r) + c^+_S V_0^\prime(r) \over 2r} 
+ { {\bf S}^{(1)}\cdot{\bf L}^{(2)}  + 
{\bf S}^{(2)}\cdot{\bf L}^{(1)}  \over m_1 m_2} {c^+_F V_2^\prime(r) \over r}
\nonumber\\
& & \!\!\!\!\!\!\!\!\! + \left( {{\bf S}^{(1)}\cdot{\bf L}^{(1)}\over m_1^2} - 
{{\bf S}^{(2)}\cdot{\bf L}^{(2)}\over m_2^2} \right) \!
{2 c^-_F V_1^\prime(r) + c^-_S V_0^\prime(r) \over 2r} 
+ { {\bf S}^{(1)}\cdot{\bf L}^{(2)}  - 
{\bf S}^{(2)}\cdot{\bf L}^{(1)}  \over m_1 m_2} {c^-_F V_2^\prime(r) \over r}
\nonumber\\
& & \!\!\!\!\!\!\!\!\! + {c_F^{(1)}c_F^{(2)}\over m_1 m_2} \left( 
{{\bf S}^{(1)}\!\cdot\!{\bf r} \, {\bf S}^{(2)}\!\cdot\!{\bf r} \over r^2} - 
{{\bf S}^{(1)}\!\cdot \!{\bf S}^{(2)} \over 3} \right) V_3(r)
\nonumber\\
& & \!\!\!\!\!\!\!\!\! +  {{\bf S}^{(1)}\!\cdot\!{\bf S}^{(2)} \over 3 m_1 m_2}
\left( c_F^{(1)} c_F^{(2)} V_4(r) -12 \, (d_{sv} +  C_F\,d_{vv}) \, \delta^3(r)\right).
\label{pot}
\end{eqnarray}
${\bf S}^{(j)} = \mbox{ {\boldmath $\sigma$}}^{(j)}/2$ and ${\bf L}^{(j)} = {\bf r}\times{\bf p}_j$ 
are the spin and orbital angular momentum operators respectively. 
The matching coefficients $d_\ell$ come from the 4-fermion sector $L_{\rm 4f}$ \cite{pineda} and 
$2 c^{\pm}_{F,S} \equiv c^{(1)}_{F,S} \pm c^{(2)}_{F,S}$. The static potential $V_0$ is 
\begin{equation}
V_0(r) = \lim_{T \to \infty}  {i\over T} \ln \langle W(\Gamma_0) \rangle, 
\quad \quad \quad
W(\Gamma) \equiv  {\rm P \,} \displaystyle\exp\left\{ -ig \oint_\Gamma dz^\mu A_\mu(z) \right\}. 
\label{v0}
\end{equation}
where $\Gamma$ is the loop made up by the quark trajectories $z_1$ and $z_2$ and the endpoints 
Schwinger strings, and the static loop $\Gamma_0$ is a $r \times T$ rectangle. The bracket means the color 
trace of the average over the gauge fields weighted by the gluon  Lagrangian  $L_{\rm g}$. 
The ``potentials" $V_1$, $V_2$, ... are scale dependent gauge field averages of electric and magnetic 
field strength insertions in the static Wilson loop, see \cite{bali,let}.  

Since the spin independent corrections come from the terms $D_0$ and ${\bf D}^2/2m$ of the Lagrangian 
(\ref{nrqcd2f}) whose matching coefficients are protected by reparameterization invariance, 
they are scale independent. An evaluation of the so-called momentum dependent corrections 
can be found in \cite{consoli}.  Terms involving logarithms of the quark masses are present. 
As a consequence of the matching procedure they are all encoded in the matching coefficients. 
Therefore, as argued in \cite{chen}, the correct handling of the matching allows finally 
the agreement between the potential derived here evaluated in the perturbative regime with 
the traditional QCD one loop perturbative calculation \cite{yndurain,ng}. For a careful discussion 
of this point see \cite{let}.

The ``potentials'' are known exactly (up to a given order in $\alpha_{\rm s}$) only in the perturbative regime, 
i.e. in the short range behaviour. Nevertheless there exist some exact relations between them which hold 
at any range. For instance, from Lorentz invariance it follows that \cite{eichten}
\begin{equation}
V^\prime_0(r) + V^\prime_1(r) - V^\prime_2(r) = 0. 
\label{grom}
\end{equation}
Since for reparameterization invariance $c^{(j)}_S = 2 c^{(j)}_F -1$ \cite{manohar}, 
equation  (\ref{grom}) holds at any scale $\mu$. Similar relations exist for the momentum dependent 
``potentials'' of Ref. \cite{consoli}. Moreover the scale independence of the potential (\ref{pot}) 
(i.e. $\displaystyle \mu {dV \over d\mu} = 0$) establishes several relations between quantities 
at different renormalization scales \cite{bali}. 

The potentials $V_0$, $V_1$, ... are suitable for direct lattice computation 
and for {\it analytic evaluation inside a QCD vacuum model}. The first possibility relies on the fact that 
all the dynamical quantities are expressed in terms of field strength insertions on a static Wilson loop 
which is an object traditionally measured on the lattice. Such an analysis has been performed in \cite{bali}. 
All the ``potentials'' have been measured on the lattice and the previous mentioned exact relations have been used 
to check the accuracy of the results. Using the parameterized form obtained in this way  for the ``potentials''  
the bound state equation has been solved and the bottomonium and charmonium spectra evaluated. The agreement 
with the experimental data is found to be quite good. On the other hand, an analytic evaluation 
of the heavy quark potential inside a QCD vacuum model turns out to be very convenient in this framework 
since by means of functional derivatives all the averages of field strength insertions on the Wilson 
loop can be expressed in terms of the average of the non-static Wilson loop alone. 
This is usually a quantity provided by QCD vacuum models. A study of different models in this 
framework has been done in \cite{brambilla,bv} where a comparison with the 
existing lattice data has also been provided. Up to now these are not accurate enough to really discriminate  
between different models. More precise lattice measurements will be performed in the near future 
providing in this way strong constraints on all the infrared QCD models with predictability on  
the long-range quarkonium interaction. 

Finally we emphasize that until soft and ultrasoft degrees of freedom have not been disentangled 
an exact value in $v$ cannot be assigned to each term of the effective 
Lagrangian and the power counting has to be interpreted at the leading order. 
For instance the $O(1)$ NRQCD Lagrangian, $L = Q_1^\dagger \left(iD_0 
+ {\bf \partial}^2 /2 m_1 \right)Q_1$ $+$ $Q_2^\dagger \left(-iD_0 + {\bf \partial}^2 /2 m_2 \right)Q_2$,  
does not contribute to (\ref{pot}) only with the static potential $V_0$. Since the 
corresponding Wilson loop ${\rm P \,} \displaystyle\exp\left\{ -ig \oint_\Gamma dz^0 A_0(z) \right\}$ 
is a function of the non-static loop $\Gamma$, its expansion produces spin independent 
terms of order $O(v)$ as well.

\section{Comments and outlook}\label{seccon}

In the framework of NRQCD we have shown how to get  the complete order $O(v^4)$ expression 
of the heavy quark-antiquark potential in terms of field strength insertions on a static Wilson loop. 
This has several advantages. Nonperturbative contributions can be evaluated by means either 
of traditional lattice calculations or of different QCD vacuum models. 
Having worked out the matching procedure, we find that the potential of Eq. (\ref{pot}) 
is consistent in the short range with the existing perturbative calculations and with 
the lattice data \cite{let}. These are sensitive to one loop and in some
cases to the next to leading correction too. We stress here that terms proportional 
to the static potential in  Eq. (\ref{pot}) have to be protected from the running.
For instance, using Eq. (\ref{grom}) and reparameterization invariance, the spin-orbit interaction 
term $\sim 2 C_F^+V_1^\prime + C_S^+V_0^\prime$ can be written as $2 C_F^+V_2^\prime - V_0^\prime$ 
where it is apparent that no Wilson coefficient multiplies the static potential. 
Moreover we can draw the following consequence about $V_2^\prime$: 
either $V_2^\prime$ does not contain any nonperturbative contribution at all 
(as present lattice data seem to suggest) or, if $V_2^\prime$ contains some nonperturbative contributions, 
they have to satisfy the RG equation $\displaystyle {d\over d\mu} C_F^+V_2^\prime = 0$. The same argument suggests 
for the Darwin term that $\Delta V_{\rm a}^E = - \Delta V_0 +$ perturbative contributions.  
This gives some constraints on the QCD vacuum models which, introducing at some point 
some approximations, lose scale invariance.  

All the corrections to the effective Lagrangian  discussed here  are relevant in order to obtain  
the quarkonium spectrum with an accuracy of $O(v^4)$. For some Wilson coefficients only the leading 
$\alpha_{\rm s}$ contribution needs to be taken into account. This is no more the case if we aim 
to reach a 10\% accuracy in the quarkonium spin splittings. Being these quantities an order $O(v^4)$ effect, 
a 10\% accuracy is achievable  only if $O(v^6)$ and $O(\alpha_{\rm s} v^4)$ effects are calculated as well. 
Therefore operators of order $O(v^6)$ should be added to the effective Lagrangian \cite{lepage}. 
The inclusion of such operators in Eq. (\ref{pot}) in terms of (two and three) field strength insertions 
on a static Wilson loop, is only a technical problem, but has not be done so far. The main reason is 
that {\it non potential contributions} are expected to become even more important. 
Ultrasoft gluonic degrees of freedom, characterized by a time scale $1/ m v^2$, exist. In perturbative QCD 
they are responsible, for instance, of the Lamb-shift (which is a $\alpha_{\rm s} v^4$ effect). 
In the language of the Wilson loop, this means that the interaction $U$ (see Sec. 3) can be 
affected by non potential contributions which have to be subtracted from $V$ in order to 
define properly a heavy quark potential. Since $\alpha_{\rm s}$ is not a small parameter at the 
ultrasoft scale, these non potential terms could be an effect of order $O(v^4)$. 
Therefore, precision studies of quarkonium have to take into account it. An approach was recently proposed in
\cite{inprep,pnrqcd}. 
The ultrasoft degrees of freedom are integrated out directly from the NRQCD Lagrangian 
giving rise to another effective theory, called potential NRQCD, where all the energy scales  
of the bound state are disentangled explicitly. The advantages are enormous, since in the new theory 
potential and non potential contributions are clearly separated. The novel  feature is that the matching 
this time takes place  in a energy region dominated  by the nonperturbative physics. 

\section*{Acknowledgements}

It is a pleasure for N. B. to thank Andrei Afanesev, Nathan Isgur, Dubravko Klabucar, Elio Soldi 
and Branko Vlahovic for the organization of this nice Conference. N. B. acknowledges financial support from  the 
European Community, 
Marie Curie fellowship-TMR Contract n. ERBFMBICT961714, and from the theory section of Milano.

\end{document}